# From climate change to pandemics: decision science can help scientists have impact

Christopher M. Baker[1,2,3], Patricia T. Campbell[4,5], Iadine Chades[6], Angela J. Dean[7,8], Susan M. Hester[3,9], Matthew H. Holden[10], James M. McCaw[1,4,5], Jodie McVernon[4,5,11], Robert Moss[5], Freya M. Shearer[5] and Hugh P. Possingham[12,13]

[1] *School of Mathematics and Statistics, The University of Melbourne, Melbourne, Australia.*

[2] *Melbourne Centre for Data Science, The University of Melbourne, Melbourne, Australia.*

[3] *Centre of Excellence for Biosecurity Risk Analysis, The University of Melbourne, Melbourne, Australia.*

[4] *Peter Doherty Institute for Infection and Immunity, The Royal Melbourne Hospital and The University of Melbourne, Australia*

[5] *Melbourne School of Population and Global Health, University of Melbourne, Parkville, Victoria 3010, Australia*

[6] *Land and Water, CSIRO, Ecosciences Precinct, Dutton Park, Queensland 4102, Australia*

[7] *Centre for the Environment, Institute for Future Environments, Queensland University of Technology, Brisbane QLD 4000*

[8] *School of Biology and Environmental Science, Queensland University of Technology, Brisbane QLD 4000*

[9] *UNE Business School, University of New England, Armidale, Australia*

[10] *Centre for Applications in Natural Resource Mathematics (CARM), University of Queensland, Australia*

[11] *Murdoch Children's Research Institute, The Royal Children's Hospital, Melbourne, Australia*

[12] *The Nature Conservancy, Arlington, VA, United States*

[13] *School of Biological Sciences, University of Queensland, Saint Lucia, Queensland 4072, Australia*

Corresponding author: Christopher M. Baker, cbaker1@unimelb.edu.au

**Abstract**

Scientific knowledge and advances are a cornerstone of modern society. They improve our understanding of the world we live in and help us navigate global challenges including emerging infectious diseases, climate change and the biodiversity crisis. For any scientist, whether they work primarily in fundamental knowledge generation or in the applied sciences, it is important to understand how science fits into a decision-making framework. Decision science is a field that aims to pinpoint evidence-based management strategies. It provides a framework for scientists to directly impact decisions or to understand how their work will fit into a decision process. Decision science is more than undertaking targeted and relevant scientific research or providing tools to assist policy makers; it is an approach to problem formulation, bringing together mathematical modelling, stakeholder values and logistical constraints to support decision making. In this paper we describe decision science, its use in different contexts, and highlight current gaps in methodology and application. The COVID-19 pandemic has thrust mathematical models into the public spotlight, but it is one of innumerable examples in which modelling informs decision making. Other examples include models of storm systems (eg. cyclones, hurricanes) and climate change. Although the decision timescale in these examples differs enormously (from hours to decades), the underlying decision science approach is common across all problems. Bridging communication gaps between different groups is one of the greatest challenges for scientists. However, by better understanding and engaging with the decision-making processes, scientists will have greater impact and make stronger contributions to important societal problems.

Scientific discoveries have fuelled enormous change in our society. Despite this, scientists are under increasing pressure to not only make discoveries and test hypotheses, but to engage with the public and contribute to solving real-world problems (Pham 2016). While improving dialogue between scientists and policy-makers is important (Cvitanovic *et al.* 2015; Berger *et al.* 2019), achieving scientific impact requires more than effective communication. For any given scientific problem, scientists need to carefully consider the wider context in which decisions are made, including differing value judgements, and the feasibility and cost of various responses. The field of decision science explicitly aims to better forecast whether actions and policies will achieve decision-makers' goals, with the ultimate aim to provide tailored, situation-specific, and relevant advice. While much scientific research is conducted without the express aim of informing policy, scientists who do want to directly influence decisions need to have a clear understanding of the utility of scientific outputs to stakeholders and the broader community. Even for those who do not aim to directly engage in that process, a sound understanding of it will help to communicate the diverse activities that constitute science.

The key feature of decision science is that it identifies, first and foremost, a *decision problem,* which includes the context of an issue. This contrasts with the more typical focus in science on either a *hypothesis* or *discovery*. Importantly, the decision problem describes a situation where some action is required, and the overall aim is to choose actions to take, rather than to learn. Learning about the system itself, while perhaps even a necessary step along the way, is not prioritised for its own sake. The decision science approach – putting objectives and actions at the forefront – is common across all decision problems (Possingham *et al.* 2001), even though the process can look very problem-specific and the actual methods used for any particular decision analysis can vary widely.

Decision science principles have been used widely in business (French *et al.* 2009), but have not yet been adopted broadly across the sciences. One area where decision science has been used widely is in spatial planning for conservation (Ball *et al.* 2009). Using decision science within the context of scientific research brings *prediction* and *optimisation* – deeply mathematical fields of science – to the fore (Guisan *et al.* 2013). But decision science is broader. Not only does it consider the specific decision problem and immediate stakeholders, but the broader social and governance implications (Figure 1). Accordingly, a decision science approach will naturally bring together a diverse multi-disciplinary team, including (at a minimum) domain experts (in, for example, the life or physical sciences), mathematical modellers and social scientists.

In this paper we describe our vision for how decision science can make an impact throughout science and in the life sciences in particular. We begin with an overview of structured decision making (Gregory *et al.* 2012) – a cornerstone decision science methodology – before describing our extended vision of decision science. We expand on the utility of decision science, describing how decision science can improve the solving of diverse real-world problems, from pandemic management plans, to large scale environmental conservation. We finish by exploring the current frontiers of decision science and identifying critical gaps.

### *Structured Decision Making*

Decision-makers are often tasked with making a meaningful contribution to solving complex problems. Structured decision making provides a framework for breaking down complex problems into manageable steps to harness the complexity. It is a participatory process that draws together objectives around an issue, determines potential actions and attempts to predict their outcomes, before finally determining a recommended pathway forward. In this section we give a brief overview of the methodology, but we encourage readers interested in using structured decision making to study more comprehensive analyses of the topic (Martin *et al.* 2009; Clemen & Reilly 2013).

*Identifying the fundamental objective(s)*

Objective-setting is an inclusive process whereby stakeholders identify the ideal end goal(s) they wish to achieve. Each specific aim should be narrowed to a set of measurable *fundamental objectives*. Importantly, the set of fundamental objectives can include multiple and even competing objectives.

*Determine potential actions*

Actions include any type of intervention that affects the system itself, such as policy changes or on-ground projects. 'No action' should always be an option we consider and doing nothing should be a conscious choice. The list of potential actions should be as diverse as possible, and not be constrained by *a priori* assumptions on pre-conceived views about feasibility and/or effectiveness.

*Predict the outcome of potential actions*

Once objectives are set, and potential actions have been identified, it is necessary to predict how well each action will help achieve the fundamental objective(s). This is often achieved through a mathematical modelling approach. Importantly, whatever approach is used to make predictions, we should quantify the uncertainty in those predictions (Milner-Gulland & Shea 2017). Scientists can make a critical contribution to this part of the structured decision making process. Scientists can provide fundamental scientific knowledge behind an uncertain process and develop models (mathematical or otherwise) to forecast the state of a system into the future in response to actions and under different scenarios.

*Determine a pathway forward*

Once the effect of implementing each action is quantified, the action, or set of actions that best meet the fundamental objective must be determined. In some situations, competing objectives may arise, where actions help achieve one objective at the expense of achieving another. In these contexts, additional processes such as multi-criteria decision analysis (Mendoza & Martins 2006) and/or conflict resolution may be required (Biggs *et al.* 2017).

Importantly, the selected actions should not be confused with the fundamental objective itself. In structured decision making, actions are never an objective that one aims to achieve from the start. However, once the best actions are identified, some refer to them as *means objectives* or *intermediate objectives* (Schwartz *et al.* 2018). But even if named as such, means objectives are but stepping stones towards achieving the fundamental objective. For example, if eliminating measles is the fundamental objective, vaccinating 95% of the population against measles may be a means objective. Where a means objective is mistaken for the fundamental objective, poor outcomes can be expected, driven first and foremost by the chance that the purpose of the action is lost, forgotten or ignored. This is like Goodhart's law: "when a measure becomes a target, it ceases to be a good measure."

*Decision Science as a pathway for impact*

Our vision of decision science for impact extends beyond structured decision making by embedding structured decision making within an overarching understanding of the social, political and organisational context in which decisions are made and implemented (Figure 1). In particular, decision analyses often focus on understanding systems and identifying actions that meet fundamental system-level objectives, while keeping economic costs low. But there are many more factors that matter, including political and logistical constraints and social and ethical accountability.

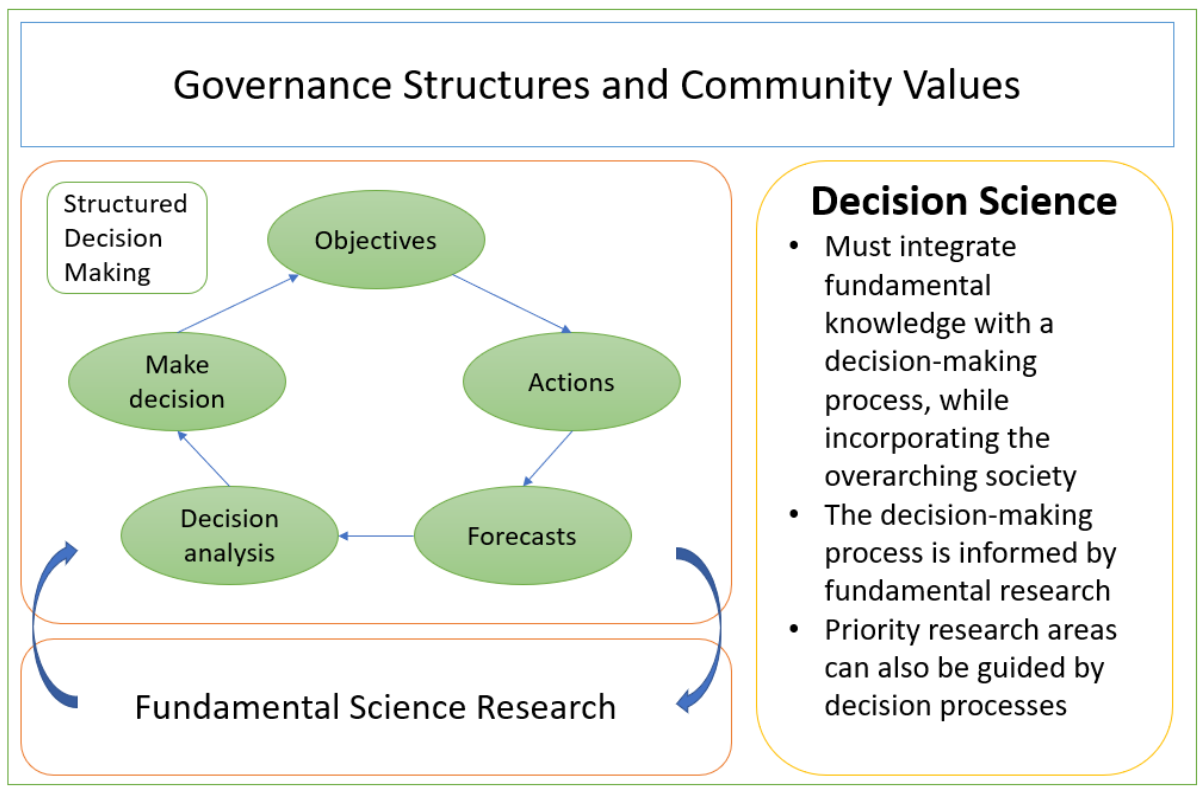

*Figure 1: Decision science must be embedded within society's overarching values and governance structures. There is a history of scientific outreach to help people and/or organisations make decisions. Embedding research within a broader discussion of stakeholders' values is a core tenet of structured decision making, which provides a coherent way to progress contentious and difficult decisions. The final layer is governance in which the broader views of government bodies and the wider community are used to inform the*

*scientific approach. This overall framework is what we define as decision science. Under this framework scientists formulate their questions in the most relevant way, with an understanding that their findings are but one part of the overall decision-making process.*

Understanding the overall context of decisions, and how scientific research fits into this puzzle, is important if scientists want their work to be used to inform decisions. The practice of integrating scientific knowledge, decision tools, stakeholder values and system governance in a single framework is what we define as decision science. As scientists interested in impact, we need to be aware of the broader context within which our work sits, even if we are not involved in all layers of that process. By understanding the decision science approach, we can make more timely and more effective contributions to evidence-based decisions.

An example of an environmental conservation approach that incorporates science, economics and decision analysis within broader value judgement and governance issues is the Project Prioritization Protocol, designed to help improve the New Zealand government's spending on threatened species (Joseph *et al.* 2009). At its core, Project Prioritization Protocol is simply a calculation of cost-effectiveness, but its key advance was integrating logistical factors, including cost and probability of success, with judgements of species value in a decision-making framework that could be operationalised by government. The Project Prioritization Protocol approach has spread internationally, including to Australia (Brazill-Boast *et al.* 2018) and the United States (Gerber 2016). The successful update of the method into government decision-making processes is due to a range of factors, including that it is: intuitive, collaborative, allows for value judgements, aligns with bureaucratic processes, and has a solid scientific and economic foundation. The protocol has not only increased efficiency, it has also increased investment (Brazill-Boast *et al.* 2018).

Of course, not all initiatives go completely to plan. Box 1 explores two case studies where the findings of the underlying science have not been fully reflected in decision outcomes. The reasons are different in each of these case studies. In the first example, a failure to acknowledge the broader context in which the science fits led to a policy failure and reduced public trust in authorities' ability to deal with environmental issues. In the second example, a failure to identify the true fundamental objective resulted in scientific advice being at odds with the policy decision.

**Box 1:** *Examples of science and decisions.*

*The 'Death-Row Dingoes': feral goat control in Australia*

Ignoring societal values can turn scientifically sound actions into public policy failures. The case of the 'death-row dingoes' in Northern Queensland, Australia in 2016 (van Eeden *et al.* 2017) is a case in point. The Hinchinbrook Shire Council released two dingoes onto Pelorus Island in a bid to cost-effectively reduce the feral goat population. Due to concerns about their impacts on local bird populations, the dingoes were implanted with poison tablets that would activate within two years to ensure they would not become a long-term problem if they evaded capture. While some would argue this was a pragmatic and scientifically sound approach to reducing the threat, it caused national outrage and many questioned the ethics of the project (Yanco *et al.* 2019). In response, the Queensland Government ordered the Council to remove the dogs, and it reduced public trust in the authorities' ability to deal with environmental issues (van Eeden *et al.* 2017).

*Risk management for bovine tuberculosis in the UK*

Public controversy can drown out a scientific solution. Bovine tuberculosis (TB) is a prominent issue in the UK, and some stakeholders advocate the widespread culling of badgers as a means to reduce transmission between herds. Bovine TB management is proving an expensive and contentious issue. The Guardian newspaper has dedicated a regular section to the issue, with over 200 articles since the year 2000 (e.g. Barkham 2015; Doward 2018). Disease transmission models "*predict that control of local badger populations and hence control of environmental transmission will have a relatively limited effect on all measures of bovine TB incidence*" (Brooks-Pollock *et al.* 2014). Despite this evidence, badger culling still continues (Carrington 2019). This example provides a cautionary tale that even when sound science is conducted, it does not mean the public policy issue will be resolved. Navigating these situations involves stepping beyond scientific analysis and requires an understanding of how to create social change.

*The role of time scales in practical decision science*

Decision science provides a framework to make decisions across a variety of contexts. Even though the fundamental approach is the same, it is important to understand how contextual factors influence the implementation of decision science. Some problems develop gradually over (e.g. climate change), while others are acute and elicit an immediate response (e.g. an Ebola outbreak or the COVID-19 pandemic). First consider climate change. While we continue to reduce our uncertainty about its future impact, intervention options and actions are already available, and, in principle, the world is both refining its understanding of the problem and implementing solutions (Wong-Parodi *et al.* 2016). In contrast, for pandemic events and natural disasters, such as earthquakes and tsunamis, the response relies on emerging knowledge gained during the event itself. While we can plan and deploy defence strategies beforehand and make systems more resilient, it is not until the event occurs that a disaster management plan can be enacted.

So, how does decision science operate for problems with such contrasting time-scales? In situations which are unfolding slowly, scientific understanding accumulates slowly, and thus we have more time for participatory decision-making processes. On the other hand, in fast time-scale problems, such as emergency response, there is minimal, if any, time for participatory decision-making processes to be conducted. Structured decision making must happen beforehand, and its findings drawn upon as the response unfolds.

*Decision Science for Pandemic Preparedness*
Pandemics are rare, fast time-scale decision problems where stakeholders must be brought together to understand objectives and actions. Preparations must happen before the start of the event, in order to establish overarching principles, and ideally, to establish the trust and information-flow arrangements that underpin good decision making. Over the past decade, governments have been largely focused on preparing for an influenza pandemic , although plans for 'Disease X' had come to some prominence in recent years (Cousins 2018). As well documented elsewhere (Tabata *et al.* 2020), a novel coronavirus, SARS-CoV-2 spread from Wuhan, China across the globe in early 2020. By late July, over 10 million infections, five hundred thousand deaths had been reported and reported case numbers were doubling every 40 days (Dong *et al.* 2020). The World Health Organization, The United States Centers for Disease Control and Prevention, and many national governments have highlighted the significant threats to society from emerging and re-emerging infectious diseases such as pandemic influenza, Ebola and SARS. As the COVID-19 pandemic has clearly demonstrated, an outbreak not only has the potential to result in many deaths, but also disrupt national and global economies. The COVID-19 pandemic has impacted virtually all international travel, disrupted global supply chains and strained medical supply logistics. Disease management approaches (many of which existed only in planning documents) have been put to the test.

The COVID-19 pandemic is a global disaster, and countries have reacted quickly, many enforcing strict measures to limit disease spread. Compared to influenza, it presents challenges due to the lack of a pharmaceutical intervention, such as antiviral drugs or a vaccine. In their absence, policy makers have relied on non-pharmaceutical interventions, including border measures and community-wide physical distancing (Chu *et al.* 2020). The early experiences in China and Italy prompted many other countries to act quickly. Closing borders, banning large gatherings and even requiring residents to stay inside their houses for weeks at a time are extreme actions, commonplace and demonstrably effective (Tanne *et al.* 2020), despite not being considered in any pre-existing pandemic management plans that we could uncover.

In this context, modelling to help inform decision-makers had to occur extremely quickly in the early stages of the pandemic (Shea *et al.* 2020). The number of publications that were released as technical reports or blog pages on institutional webpages and then as open access pre-prints has been enormous, surpassing any other topic in terms of rate of increase, and fortuitously, modelling of infectious diseases is a well-established field (Anderson & May 1979) and there has been a recent focus on incorporating models and decision analysis into pandemic planning (Alahmadi *et al.* 2020; Shearer *et al.* 2020). As both the decision-making process and the scientific knowledge of COVID-19 are moving quickly, scientists seeking to contribute to policy arguably need to already have trusted relationships with decision-makers in place to enable them to provide relevant scientific advice at the right time. Despite the short timeframes, the process of decision science and structured decision making remains the same. They key difference is that with less time, there are fewer opportunities to resolve uncertainty and make precise predictions. But this is a typical situation for an adaptive management cycle (Chadès *et al.* 2016), where decisions are made, data is collected, predictions are improved, and decisions are refined in a cycle.

Decision making at the beginning of the COVID-19 pandemic focused on immediate health impacts of the virus and necessarily moved quickly, but as the pandemic progresses, the need to move from a fast decision problem, to a medium to long-term decision problem is becoming clear. While country lockdowns have been extremely successful in reducing transmission, they come with a raft of other costs, both economic, social and related to (non-COVID) health issues. Response options in this situation – faced the world over – constitute a challenging decision problem. It involves multiple competing objectives, occurring over multiple time-scales. These objectives will be valued differently by different people. And predictions of possible actions will all likely have large uncertainties due to the unprecedented nature of the event – both in terms of modelling the transmission of the virus, but also modelling the socio-economic outcomes of alternative responses given that the system under study has been impacted so strongly that many assumptions underlying socio-economic models may reasonably be challenged. On top of those complexities, how members of society respond to policies – when they have been subject to unprecedented and unexpected disruption and trauma, is both highly uncertain and likely highly heterogeneous. Decision science has approaches to address each of these challenges: when there are multiple objectives, multi-criteria decision analysis is useful (Marttunen *et al.* 2017), when there is uncertainty, value of information theory can suggest the most useful data to collect (Canessa *et al.* 2015), and incentive theory and behavioural sciences can help predict people's response to policy change (Laffont & Martimort 2009; Bavel *et al.* 2020).

Beyond differing values around population health and the economy, there are also fundamental decisions to be made about whether elimination or suppression is the aim in responding to COVID-19 (although, a third strategy is to aim for herd immunity). While elimination within a region may allow resumption of many daily activities, it leaves the population susceptible and at high risk of future incursions. The other broad option is suppression, where measures would be kept in place for longer timeframes, with the goal of keeping prevalence stable and sufficiently low to ensure public health and clinical response remain effective. Suppression means that any outbreaks, or fluctuations in incidence, would be managed and die out, giving the population long-term protection, even if there are overseas introductions. Choosing between these options is not about value judgements. Rather it is about risk preference. The elimination option could have tremendous payoff, but comes at a higher risk of failure, and potentially delivers a 'false sense of security' to the population, itself a risk given the risk of failure. Robust decision making is the idea of choosing actions that minimise the probability of a bad outcome (Moore & Runge 2012). A risk-averse decision-maker would likely choose suppression (minimise probability of worst outcome), while another may maximise the chance of a good outcome (maximise the expected outcome).

Decision science provides a way to explicitly frame scientific questions about COVID-19 within the relevant community and government structures. Community consultation is required to properly understand the values of a population, and clear communication is important to ensure that everyone can understand the fundamentals of disease transmission and the potential outcomes of different management strategies. While these aspects bring together governance, community engagement and science, we must also include social and behavioural science, as it is critical to understand how policy influences community behaviours and therefore population transmission dynamics. Underlying all of these themes is trust, which must exist between scientists, governments and the general population. Losing public trust can undermine progress ("Building trust is essential to combat the Ebola outbreak" 2019; Bavel *et al.* 2020).

**Building relationships for effective decision science**

Decision science facilitates effective dialogue and relationships between scientists and decision-makers by providing a framework that requires scientists to elicit decision-maker opinions and values. Firstly, it demands that the scientists understand the decision context. A better understanding of decision-maker needs—type of inputs needed, the local context, the timing of decisions, stakeholder preferences and values, the degree of risk or uncertainty that is acceptable—can enable tailoring of scientific outputs for best application

(Patt 2009; Marcot *et al.* 2012). Secondly, dialogue is also pivotal for fostering relationships, shared understanding, trust (from the decision-maker) in the scientific integrity of the process, and trust (from the scientist) that cautions and uncertainties in predictions will be considered appropriately (Patt 2009; Winterfeldt 2013; Doyle *et al.* 2014). Dialogue also allows scientists to gauge how their information is received, and provides them with the opportunity to clarify the scientific process, explain potential caveats or ways to apply findings across different contexts (Patt 2009).

One of the specific challenges in decision science is how to communicate uncertainty in a way that supports decision making rather than inadvertently eliciting indecision or rejection of scientific information. Effective communication of uncertainty can build trust, and conversely, avoiding communication of uncertainty can undermine it (Patt 2009). Uncertainty can arise from a range of sources: incomplete knowledge of all parameters within a complex system, incomplete control of how a response will be implemented, and errors in measurement of estimates (Johnson *et al.* 2015). The style of communication can also inadvertently contribute to uncertainty. For example, numerical likelihoods (e.g. '75% chance of occurring') are perceived to be more certain than descriptive likelihoods (e.g. 'Likely to occur') which are often misinterpreted (Doyle *et al.* 2014). In communicating uncertainty, scientists must reduce uncertainties to their 'decision-relevant' elements, describing the source of uncertainty, and incorporating uncertainty into estimates via use of credible intervals or ranges (Fischhoff & Davis 2014). Ranges and credible intervals may incorporate a range of sources of uncertainty, including variability in estimates' applicability to the decision context, and overall strength of the science (Fischhoff & Davis 2014).

**The Future of Decision Science**

So how, as scientists, can we reframe our practice to maximise our contribution to public policy? First and foremost, we must understand the context in which the technical work we perform might be used. Particularly for those who are impact-focused, technical work should be embedded within the decision science framework, acknowledging the societal values and political context in which decisions are made. If as a scientific community more of us can engage and work in this way, science will play a greater and more effective role in public policy. Secondly, all scientists must recognise that resolving uncertainty is not always necessary for decision making. Some unknowns are critical to resolve, while others are irrelevant for decisions (Li *et al.* 2017). Critical aspects of uncertainty in the context of decision making are the objectives and the list of possible actions; and changing the actions or objectives alters which system uncertainties are important.

While the decision science toolkit is well-developed, there are still areas where our current approaches are incomplete, which include balancing complexity and interpretability of solutions, and where stakeholders have antithetical aims and values. Complex decision problems often demand complex science and modelling, but increasing complexity makes it hard to understand and interpret results (Ferrer-Mestres *et al.* 2020). While creating tools that stakeholders can interact with helps understanding and dialogue (Winterfeldt 2013), we should aim to develop simple tools to aid decision-makers, that still convey the complexity and subtlety of the situation. Decision scientists have developed methods to deal with multiple competing objectives (Mendoza & Martins 2006; Marttunen *et al.* 2017), but these methods are designed for situations where there are trade-offs (for example, between biodiversity and economic gain), and are not equipped for situations where stakeholders are unwilling to make any trade-offs. The illegal wildlife trade is a prime example where there are extreme differences in opinion (Biggs *et al.* 2017) and decision science does not have an effective way forward.

Throughout this paper we have articulated how decision science fits within public policy formation and how scientists can better align their work to increase its influence. Many of the topics are complex and are research areas in their own right, and continued research will improve our ability to engage in complex decision problems. However, despite the broad utility of decision science, we consider the name to be a misnomer. It's less of a science than a problem-solving approach, and there's no single 'recipe for success'. The point of decision science is to view scientific problems from a decision-making perspective, focusing on objectives and feasible actions to identify critical scientific questions that will inform decisions. As such, the more widely the tenets of decision science are understood, there can be better science-based decisions and more real-world scientific impact.

**Acknowledgements**

We acknowledge funding from The ARC Centre of Excellence for Environmental Decisions, The NHMRC Centres of Research Excellence *Policy Relevant Infectious Disease Simulation and Mathematical Modelling* (PRISM2) and *Supporting Participatory Evidence use for the Control of Transmissible diseases in our Region Using Modelling* (SPECTRUM), The ARC Centre of Excellence for Biodiversity Risk Analysis and The ARC Centre of Excellence for Mathematical and Statistical Frontiers.